\documentclass[%
 reprint,
 amsmath,amssymb,
 aps,
 prd,
 noeprint,
 superscriptaddress,
]{revtex4-2}

\usepackage{graphicx}
\usepackage{dcolumn}
\usepackage{bm}
\usepackage{comment}
\usepackage{appendix}
\usepackage{diagbox}
\usepackage{hyperref}

\begin{document}
\title{Longitudinal double-spin asymmetry for inclusive jet and dijet production in polarized proton collisions at $\sqrt{s}=200\,$GeV}
\affiliation{Abilene Christian University, Abilene, Texas   79699}
\affiliation{AGH University of Science and Technology, FPACS, Cracow 30-059, Poland}
\affiliation{Alikhanov Institute for Theoretical and Experimental Physics NRC "Kurchatov Institute", Moscow 117218, Russia}
\affiliation{Argonne National Laboratory, Argonne, Illinois 60439}
\affiliation{American University of Cairo, New Cairo 11835, New Cairo, Egypt}
\affiliation{Brookhaven National Laboratory, Upton, New York 11973}
\affiliation{University of California, Berkeley, California 94720}
\affiliation{University of California, Davis, California 95616}
\affiliation{University of California, Los Angeles, California 90095}
\affiliation{University of California, Riverside, California 92521}
\affiliation{Central China Normal University, Wuhan, Hubei 430079 }
\affiliation{University of Illinois at Chicago, Chicago, Illinois 60607}
\affiliation{Creighton University, Omaha, Nebraska 68178}
\affiliation{Czech Technical University in Prague, FNSPE, Prague 115 19, Czech Republic}
\affiliation{Technische Universit\"at Darmstadt, Darmstadt 64289, Germany}
\affiliation{ELTE E\"otv\"os Lor\'and University, Budapest, Hungary H-1117}
\affiliation{Frankfurt Institute for Advanced Studies FIAS, Frankfurt 60438, Germany}
\affiliation{Fudan University, Shanghai, 200433 }
\affiliation{University of Heidelberg, Heidelberg 69120, Germany }
\affiliation{University of Houston, Houston, Texas 77204}
\affiliation{Huzhou University, Huzhou, Zhejiang  313000}
\affiliation{Indian Institute of Science Education and Research (IISER), Berhampur 760010 , India}
\affiliation{Indian Institute of Science Education and Research (IISER) Tirupati, Tirupati 517507, India}
\affiliation{Indian Institute Technology, Patna, Bihar 801106, India}
\affiliation{Indiana University, Bloomington, Indiana 47408}
\affiliation{Institute of Modern Physics, Chinese Academy of Sciences, Lanzhou, Gansu 730000 }
\affiliation{University of Jammu, Jammu 180001, India}
\affiliation{Joint Institute for Nuclear Research, Dubna 141 980, Russia}
\affiliation{Kent State University, Kent, Ohio 44242}
\affiliation{University of Kentucky, Lexington, Kentucky 40506-0055}
\affiliation{Lawrence Berkeley National Laboratory, Berkeley, California 94720}
\affiliation{Lehigh University, Bethlehem, Pennsylvania 18015}
\affiliation{Max-Planck-Institut f\"ur Physik, Munich 80805, Germany}
\affiliation{Michigan State University, East Lansing, Michigan 48824}
\affiliation{National Research Nuclear University MEPhI, Moscow 115409, Russia}
\affiliation{National Institute of Science Education and Research, HBNI, Jatni 752050, India}
\affiliation{National Cheng Kung University, Tainan 70101 }
\affiliation{Nuclear Physics Institute of the CAS, Rez 250 68, Czech Republic}
\affiliation{Ohio State University, Columbus, Ohio 43210}
\affiliation{Institute of Nuclear Physics PAN, Cracow 31-342, Poland}
\affiliation{Panjab University, Chandigarh 160014, India}
\affiliation{Pennsylvania State University, University Park, Pennsylvania 16802}
\affiliation{NRC "Kurchatov Institute", Institute of High Energy Physics, Protvino 142281, Russia}
\affiliation{Purdue University, West Lafayette, Indiana 47907}
\affiliation{Rice University, Houston, Texas 77251}
\affiliation{Rutgers University, Piscataway, New Jersey 08854}
\affiliation{Universidade de S\~ao Paulo, S\~ao Paulo, Brazil 05314-970}
\affiliation{University of Science and Technology of China, Hefei, Anhui 230026}
\affiliation{Shandong University, Qingdao, Shandong 266237}
\affiliation{Shanghai Institute of Applied Physics, Chinese Academy of Sciences, Shanghai 201800}
\affiliation{Southern Connecticut State University, New Haven, Connecticut 06515}
\affiliation{State University of New York, Stony Brook, New York 11794}
\affiliation{Instituto de Alta Investigaci\'on, Universidad de Tarapac\'a, Arica 1000000, Chile}
\affiliation{Temple University, Philadelphia, Pennsylvania 19122}
\affiliation{Texas A\&M University, College Station, Texas 77843}
\affiliation{University of Texas, Austin, Texas 78712}
\affiliation{Tsinghua University, Beijing 100084}
\affiliation{University of Tsukuba, Tsukuba, Ibaraki 305-8571, Japan}
\affiliation{United States Naval Academy, Annapolis, Maryland 21402}
\affiliation{Valparaiso University, Valparaiso, Indiana 46383}
\affiliation{Variable Energy Cyclotron Centre, Kolkata 700064, India}
\affiliation{Warsaw University of Technology, Warsaw 00-661, Poland}
\affiliation{Wayne State University, Detroit, Michigan 48201}
\affiliation{Yale University, New Haven, Connecticut 06520}

\author{M.~S.~Abdallah}\affiliation{American University of Cairo, New Cairo 11835, New Cairo, Egypt}
\author{J.~Adam}\affiliation{Brookhaven National Laboratory, Upton, New York 11973}
\author{L.~Adamczyk}\affiliation{AGH University of Science and Technology, FPACS, Cracow 30-059, Poland}
\author{J.~R.~Adams}\affiliation{Ohio State University, Columbus, Ohio 43210}
\author{J.~K.~Adkins}\affiliation{University of Kentucky, Lexington, Kentucky 40506-0055}
\author{G.~Agakishiev}\affiliation{Joint Institute for Nuclear Research, Dubna 141 980, Russia}
\author{I.~Aggarwal}\affiliation{Panjab University, Chandigarh 160014, India}
\author{M.~M.~Aggarwal}\affiliation{Panjab University, Chandigarh 160014, India}
\author{Z.~Ahammed}\affiliation{Variable Energy Cyclotron Centre, Kolkata 700064, India}
\author{I.~Alekseev}\affiliation{Alikhanov Institute for Theoretical and Experimental Physics NRC "Kurchatov Institute", Moscow 117218, Russia}\affiliation{National Research Nuclear University MEPhI, Moscow 115409, Russia}
% ...

\collaboration{STAR Collaboration}\noaffiliation


\begin{abstract}
We report high-precision measurements of the longitudinal double-spin asymmetry, $A_{LL}$, for midrapidity inclusive jet and dijet production in polarized $pp$ collisions at a center-of-mass energy of $\sqrt{s}=200\,\mathrm{GeV}$. The new inclusive jet data are sensitive to the gluon helicity distribution, $\Delta g(x,Q^2)$, for gluon momentum fractions in the range from $x \simeq 0.05$ to $x \simeq 0.5$, while the new dijet data provide further constraints on the $x$ dependence of $\Delta g(x,Q^2)$. The results are in good agreement with previous measurements at $\sqrt{s}=200\,\mathrm{GeV}$ and with recent theoretical evaluations of prior world data. Our new results have better precision and thus strengthen the evidence that $\Delta g(x,Q^2)$ is positive for $x > 0.05$.
\end{abstract}

\maketitle

The origin of the spin of the nucleon in terms of its constituent quark, antiquark, and gluon spins and their orbital angular momenta is a fundamental challenge for strong interaction physics. Polarized deep-inelastic scattering (DIS) experiments~\cite{Alguard:1978gf,Baum:1983ha,Ashman:1989ig,Adeva:1998vv,Anthony:1996mw,Abe:1998wq,Abe:1997cx,Anthony:1999rm,Anthony:2000fn,Ackerstaff:1997ws,Airapetian:2007mh,Alekseev:2010hc,Adolph:2015saz,Adolph:2016myg,Abe:1997cx, Abe:1997cx,Parno:2014xzb,Prok:2014ltt,Guler:2015hsw} have shown that less than a third of the nucleon spin originates from the spins of quarks and antiquarks~\cite{deFlorian:2014yva, deFlorian:2019zkl, Nocera:2014gqa, Leader:2014uua, Sato:2016tuz, Ethier:2017zbq, Khanpour:2017cha, Khorramian:2020gkr, Bourrely:2014uha, Lin:2017snn, Salimi-Amiri:2018had, deFlorian:2019egz}.
Semi-inclusive polarized DIS experiments~\cite{Ashman:1989ig,Adeva:1997qz, Ackerstaff:1999ey, Airapetian:2004zf, Alekseev:2009ac, Alekseev:2010ub} and measurements of polarized hadroproduction of $W$-bosons~\cite{Adare:2010xa,Aggarwal:2010vc, Adamczyk:2014xyw,Adare:2015gsd,Adare:2018csm,Adam:2018bam} have delineated the quark and antiquark spin contributions by flavor and have recently revealed an asymmetry in the polarized light quarksea~\cite{Adam:2018bam}.
Measurements of the spin-dependent rates of jets \cite{Abelev:2006uq,Abelev:2007vt,Adamczyk:2012qj,Adamczyk:2014ozi,Adam:2019aml}, dijets \cite{Adamczyk:2016okk,Adam:2018pns,Adam:2019aml}, and $\pi^0\mathrm{s}$ \cite{Adare:2007dg,Adare:2008aa,Adare:2008qb,Abelev:2009pb,Adamczyk:2013yvv,Adare:2014hsq,Adare:2015ozj,Adam:2018cto} produced in
polarized $pp$ collisions at RHIC provide evidence for positive gluon polarization with a strong constraint from the jet data at a center-of-mass energy of $\sqrt{s} = 200\,\mathrm{GeV}$~\cite{deFlorian:2014yva,Nocera:2014gqa}. Perturbative QCD analyses~\cite{deFlorian:2014yva, deFlorian:2019zkl,Nocera:2014gqa} of the world data at next-to-leading order (NLO) precision suggest that gluon spins could contribute $\simeq\!40\%$ to the spin of the proton for gluon fractional momenta $x > 0.05$ at a scale of $Q^2 = 10\,(\mathrm{GeV}/c)^2$. The corresponding RHIC spin-averaged differential production cross sections~\cite{Abelev:2006uq,Abelev:2009pb,Adamczyk:2013yvv,Adamczyk:2016okk,Adare:2007dg} are well described at NLO.

In this paper, we present new measurements of the double-spin asymmetry,
\begin{equation}
    A_{LL} \equiv \frac{\sigma^{++}-\sigma^{+-}}{\sigma^{++}+\sigma^{+-}},
\end{equation}
for inclusive jets and dijets produced in longitudinally polarized proton-proton collisions at $\sqrt{s} = 200\,\mathrm{GeV}$.
Here, $\sigma^{++}$ ($\sigma^{+-}$) denotes the jet or dijet differential production cross section when the colliding protons have equal (opposite) helicities. The data were recorded by the STAR experiment in the year 2015 and correspond to an integrated luminosity of $\mathcal{L} = 52\,\mathrm{pb}^{-1}$.
Data for different beam spin configurations were recorded in short succession by injecting beam bunches with a pattern of proton polarizations into RHIC and by changing this pattern between beam fills.
Spin rotator magnets upstream and downstream of the STAR interaction region rotated the proton beam spins from and to the stable vertical direction in RHIC to provide longitudinal spin directions at STAR.
The luminosity-weighted product of polarizations, $P$, for the two beams was $P^2 = 0.30$, 
measured with a relative uncertainty of 6.1\%~\cite{schmidke:rhic-polarization}. The values were obtained from \textit{in situ} measurements of the individual RHIC beams with proton-carbon polarimeters~\cite{Jinnouchi:2004up} that were calibrated using a polarized atomic hydrogen gas jet target~\cite{OKADA2006450}. This data set has a figure of merit, $P^{4}\mathcal{L}$, that is about 2 times larger than that of our previous measurements at 200\,GeV  \cite{Adamczyk:2014ozi,Adamczyk:2016okk,Adam:2018pns}. 

The STAR detector subsystems used in these measurements are the Time Projection Chamber (TPC) \cite{Anderson:2003ur} and the Barrel (BEMC) \cite{Beddo:2002zx} and Endcap Electromagnetic Calorimeters (EEMC) \cite{Allgower:2002zy}. The TPC measures charged particle trajectories in a 0.5 T solenoidal magnetic field in the pseudorapidity range $|\eta| < 1.3$ for all azimuthal angles. The BEMC and EEMC cover $|\eta| < 1$ and $1.1 < \eta < 2$, respectively, for all azimuthal angles. Jet patch (JP) triggers requiring transverse energy deposits in fixed $\Delta\eta \times \Delta\phi =  1 \times 1$ regions of the BEMC or EEMC in excess of $5.4\,\mathrm{GeV}$ for the JP1 trigger and $7.3\,\mathrm{GeV}$ for the JP2 trigger were used to initiate the experiment readout. The JP1 trigger was prescaled by a factor of about 10, while there was no prescale for JP2. The vertex position detectors (VPDs) \cite{Llope:2014nva} and the zero degree calorimeters (ZDCs) \cite{Adler:2000bd}, covering forward $\eta$ regions of $4.2 < |\eta| < 5.2$ and $|\eta| > 6.6$, respectively, were used to determine the relative luminosities for different helicity states of the colliding beams. The ZDCs are equipped with segmented shower maximum detectors and were also used to measure residual transverse beam polarization components at STAR. These components were smaller than 10\% of the total polarization.

The analysis methods were similar to those in the most recent STAR inclusive jet and dijet $A_{LL}$ analyses \cite{Adam:2019aml}. Jets were reconstructed from the tracks measured by the TPC and the energy deposits in the BEMC and EEMC towers. The anti-$k_T$ algorithm \cite{Cacciari:2008gp}, as implemented in the FastJet 3.0.6 package \cite{Cacciari:2011ma}, was used with the same resolution parameter, $R = 0.6$, as for our previous $200\,\mathrm{GeV}$ analyses \cite{Adamczyk:2014ozi,Adamczyk:2016okk,Adam:2018pns}.

Tracks were required to have a transverse momentum of $p_T \geq 0.2\,\textrm{GeV}/c$. BEMC and EEMC towers were required to have a transverse energy deposit of $E_T \geq 0.2\,\textrm{GeV}$. Additionally, tracks were required to have at least 12 hit points in the TPC, and have registered at least 51\% of the possible hits along the reconstructed track segment. 
As in Ref.~\cite{Adam:2019aml}, tracks were required to be associated with the collision vertex for the event by imposing a $p_T$-dependent distance of closest approach cut and a correction  was applied to the BEMC or EEMC tower $E_T$ if a track pointed to the tower.

Inclusive jet $p_T$ and dijet invariant mass, $M_{inv}$, were corrected for underlying-event contributions using the off-axis technique of Ref.~\cite{ALICE:2014dla} as adapted in Ref.~\cite{Adam:2019aml}. For each jet, the TPC tracks and calorimeter hits that fell within two off-axis cones with radius $R = 0.6$, at the same $\eta$ as the jet but $\pm \pi/2$ away in $\phi$, were selected. The information from these off-axis cones was used to subtract contributions from the underlying event on a jet-by-jet basis.

Jets were divided into mutually exclusive categories based on the highest jet patch trigger that they satisfied.
In the inclusive-jet analysis, only jets that pointed toward a triggered jet patch were considered. The required minimum jet $p_T$ after the underlying-event correction was $6.0\,\textrm{GeV}/c$ for jets reconstructed from the JP1 sample, and $8.4\,\textrm{GeV}/c$ for the JP2 sample. The jet axis was required to have $|\eta|<1$ and the analysis was performed in three $\eta$ intervals.
Jets that included a track with $p_T>30\,\textrm{GeV}/c$ were rejected because these tracks often had poor resolution.
To suppress beam gas and cosmic ray backgrounds, the neutral energy fraction in the jet was required to be smaller than 0.95 and the summed $p_T$ of charged-particle tracks was required to be larger than $0.5\,\textrm{GeV}/c$.
Moreover, a small fraction of low-$p_T$ jets were rejected since the underlying-event correction would shift jet $p_T$ by more than two jet-$p_T$ intervals.
In about 4\% of the inclusive-jet events, the event had two jets that satisfied all analysis selection criteria, and fewer than 0.03\% of the events contained more jets. For these events, the two jets with the highest $p_T$ were further analyzed.

In the dijet analysis, candidates were selected from the two highest $p_T$ jets in the event without imposing the inclusive jet selection criteria. The jet axes in the candidate dijet event were required to be more than 120$^{\circ}$ apart in azimuth and within $|\eta|<0.8$. At least one jet in the pair was required to contain energy from charged tracks.
For dijets that contained a track with $p_T > 30\,\mathrm{GeV}/c$, the reconstructed jet-$p_T$ values were required to agree to within 50\% to ensure that the transverse momenta of the two jets were balanced. Dijets failing this requirement arise from poorly reconstructed tracks with artificially high $p_T$ and were rejected from the sample.
The underlying-event correction on the dijet $M_{inv}$ was required to be less than 36\% of the reconstructed $M_{inv}$, resulting in most dijet events being shifted by no more than one dijet-$M_{inv}$ interval. To enable comparisons with theoretical predictions, an asymmetric transverse momentum selection criterion was applied to the jets, such that one jet had $p_T \geq 8\,\textrm{GeV}/c$ and the other had $ p_T \geq 6\,\textrm{GeV}/c$ \cite{deFlorian:1998qp}. Last, at least one jet in the dijet pair was required to point toward a triggered jet patch. If either jet was categorized as JP2, then the dijet event was classified as a JP2 dijet event. Otherwise, provided at least one of the jets was categorized as JP1, the dijet event was classified as a JP1 dijet event. Dijet events were categorized into $\textrm{sign}(\eta_1) = \textrm{sign}(\eta_2)$ and $\textrm{sign}(\eta_1) \neq \textrm{sign}(\eta_2)$ topologies, where $\eta_1$ and $\eta_2$ are the pseudorapidities of the two jets. The different dijet event topologies sample different regions of $x$ \cite{Adamczyk:2016okk}. 

Simulated events were used to correct the reconstructed jet quantities for detector response and to estimate contributions to the systematic uncertainties. In these analyses, $pp$ events were generated using the PYTHIA 6.4.28 \cite{Sjostrand:2006za} event generator with the Perugia 12 \cite{Skands:2010ak} tune. In addition, the PARP(90) parameter controlling the energy dependence of the low-$p_T$ cut off for the underlying-event generation process was reduced  as
in Ref.~\cite{Adam:2019aml} to ensure that the inclusive $\pi^{\pm}$ yields for $p_T<3\,\textrm{GeV}/c$ from the STAR measurements at $\sqrt{s}=200\,\textrm{GeV}$ \cite{Adams:2006nd, Agakishiev:2011dc} were better reproduced. The generated events were processed through a STAR
detector-response package based on GEANT 3 \cite{Brun:1994aa} and then embedded into zero-bias events to ensure the same beam background and pile-up contributions as in the data.

Jets were reconstructed from the simulated charged-particle tracks in the TPC and calorimeter responses on the detector level. Parton-level jets were reconstructed from the hard-scattered partons in the collision, including those from initial- and final-state radiation, but not those from the underlying event and beam remnants.

\begin{figure}
\includegraphics[width=\columnwidth]{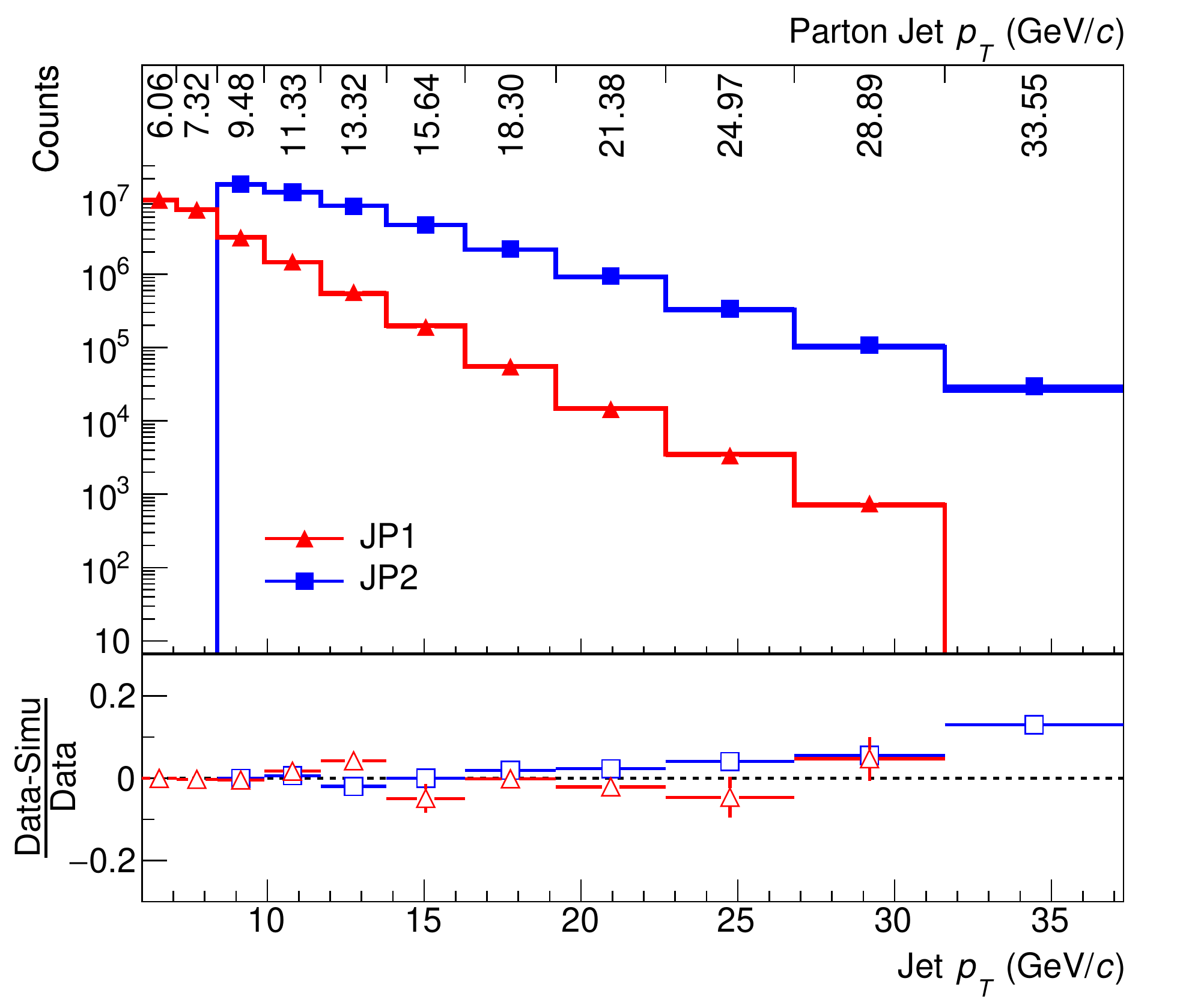}
\caption{\label{fig:JetPt} Inclusive jet yield for $|\eta| < 1$ versus jet $p_{T}$ at the detector level (lower scale) and at the parton level (upper scale). The upper panel shows the data for the JP1 and JP2 trigger conditions as points and the corresponding simulations as histograms. The lower panel shows the relative differences between data and simulation with their statistical uncertainties as vertical error bars, unless they are smaller than the marker size.}
\end{figure}

\begin{figure}
\includegraphics[width=\columnwidth]{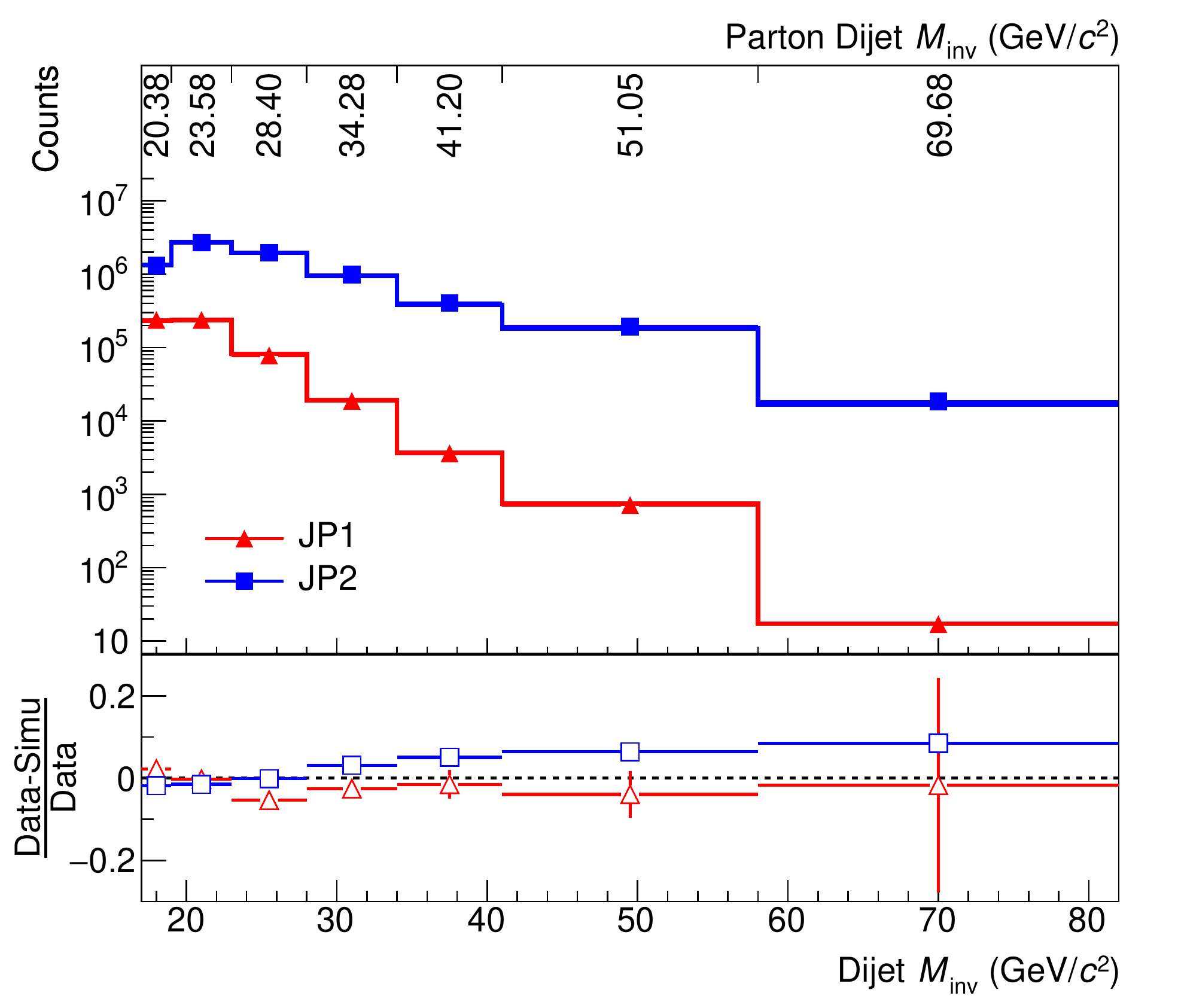}
\caption{\label{fig:DijetMass} Dijet yield versus the dijet $M_{inv}$ at the detector level (lower scale) and at the parton level (upper scale). The upper panel shows the data for the JP1 and JP2 trigger conditions as points and the corresponding simulations as histograms. The lower panel shows the relative differences between data and simulation with their statistical uncertainties as vertical error bars, unless they are smaller than the marker size.}
\end{figure}

To compare our results with theoretical predictions at the parton level, a correction was applied to
the reconstructed jet $p_T$ (dijet $M_{inv}$) in every jet-$p_T$ (dijet-$M_{inv}$) interval.
This shift was determined by comparing detector-level and their corresponding parton-level jets. Figure~\ref{fig:JetPt} compares the data and embedded simulations for JP1 and JP2 jet yield versus jet $p_{T}$ at the detector level (lower scale) and the parton level (upper scale).
The dijet yield versus dijet $M_{inv}$ is shown in Fig.~\ref{fig:DijetMass}.
The data and embedded simulation agree to within 13\%.
The effects from differences at this level are covered by the systematic uncertainties.

The values of $A_{LL}$ were evaluated according to
\begin{equation}
    A_{LL} = \frac{\sum_\mathrm{runs}P_BP_Y(N^{++}-rN^{+-})}{\sum_\mathrm{runs}P_B^2P_Y^2(N^{++}+rN^{+-})},
\end{equation}
where $P_B$ and $P_Y$ are the measured polarizations of the RHIC ``blue'' ($B$) and ``yellow'' ($Y$) beams~\cite{schmidke:rhic-polarization}, $N^{++}$ and $N^{+-}$ are the yields from colliding beam bunches where the protons preferentially had equal or opposite helicities, respectively,
and $r$ is the relative luminosity for collisions with these helicity configurations. The relative luminosity and beam polarizations were determined for each experiment run, typically 30 minutes in duration. The average beam polarizations for the analyzed data are $\left<P_B\right> = 0.523 \pm 0.016$ and $\left<P_Y\right> = 0.565 \pm 0.017$.

The values of $r$ were determined using scalers that recorded single and coincident counts from the VPD stations on either side of the interaction region for the different beam spin configurations for each run.
The observed event counts were corrected for accidental and multiple coincidences using the method of Ref.~\cite{CroninHennessy:1999hf}, and the corrected yields were summed for each spin combination. The corrected yields for each run were then used to determine $r$ for that run.
The relative luminosity has a multimodal distribution ranging between 0.96 and 1.04, depending on the polarization pattern and intensities of the injected beam bunches, and an average value of $\left<r\right> = 1.003$ for the analyzed data.
The systematic uncertainty in $r$ was estimated to be $0.0005$ by comparing results from the VPD and ZDC. This contributes $0.0007$ to the systematic uncertainty for each $A_{LL}$ value.

The asymmetry $A_{LL}$ contains leading order contributions from quark-quark, quark-gluon, and gluon-gluon scattering processes. Differences in the trigger efficiencies for different subprocesses could introduce a bias in $A_{LL}$, as could distortions due to the finite resolution of the detector. To account for these, a trigger bias correction was determined following the same procedure as in Ref.~\cite{Adam:2019aml}. This method utilizes the embedded simulation sample and 100 equally probable replicas from NNPDFpol1.1 \cite{Nocera:2014gqa} to calculate $A_{LL}$ using PYTHIA parton kinematics from the simulation for both parton-level and detector-level jets (dijets). The difference between parton-level and detector-level $A_{LL}$ was determined for each replica and for each $p_T$ ($M_{inv}$) interval separately.
The mean of the differences was used as the trigger bias correction.
The magnitude of this correction is at most 0.0016 (0.0092) in the inclusive jet (dijet) analysis.
The associated uncertainty ranges from 0.0002 (0.0005) to 0.0012 (0.0033), exhibiting upward trends with increasing $p_T$ ($M_{inv}$).

The underlying event can increase the apparent energy in the jets. To assess the associated uncertainty contribution to $A_{LL}$, we calculated the change in the cross section for jets (dijets) that would occur if the $p_T$ ($M_{inv}$) intervals were shifted by the product of the mean underlying-event contribution, $dp_T$ ($dM_{inv}$), and the asymmetry of the underlying-event contribution, $A_{LL}^{dp_T}$ ($A_{LL}^{dM_{inv}}$), defined as
\begin{equation}
    A_{LL}^{dp_T} = \frac{1}{P_{B}P_{Y}}\frac{\langle dp_T\rangle^{++} - \langle dp_T\rangle^{+-}}{\langle dp_T\rangle^{++} + \langle dp_T\rangle^{+-}}.
\end{equation}
Here, $\langle dp_T\rangle^{++}$ ($\langle dp_T\rangle^{+-}$) is the average underlying-event correction for the same (opposite) beam helicity combination. There is a corresponding definition for $A_{LL}^{dM_{inv}}$. 
A constant fit gives $A_{LL}^{dp_T} = 0.0006 \, \pm \, 0.0006$ for $\eta$ region $0.5<|\eta|<1$, and $A_{LL}^{dp_T} = 0.0021 \, \pm \, 0.0004$ for $|\eta| < 0.5$. For dijets $A_{LL}^{dM_{inv}} = 0.0014 \, \pm \, 0.0013$  for the $\textrm{sign}(\eta_1) = \textrm{sign}(\eta_2)$ topology, and $A_{LL}^{dM_{inv}} = -0.0004 \, \pm \, 0.0012$ for the $\textrm{sign}(\eta_1) \neq \textrm{sign}(\eta_2)$ topology. 
The corresponding uncertainty is one of the leading contributors to the $A_{LL}$ systematic uncertainty for both inclusive jets and dijets. For the inclusive jets (dijets) the uncertainty is $0.0010$ ($0.0007$) for the lowest $p_T$ ($M_{inv}$) interval, and decreases with increasing  $p_T$ ($M_{inv}$).
The uncertainty on the jet $p_T$ (dijet $M_{inv}$) associated with the underlying-event subtraction was determined by comparing the mean correction, $dp_T \approx 0.7 \,\mathrm{GeV}/c$ ($dM_{inv} \approx 1.3\, \mathrm{GeV}/c^2$),
in data to the one from embedded simulation.
The difference for each $p_T$ ($M_{inv}$) interval was assigned as the systematic uncertainty.

An uncertainty on the corrected jet $p_T$ (dijet $M_{inv}$) due to the PYTHIA tune was evaluated by using several variants of the Perugia 12 tune as in Ref.~\cite{Adam:2019aml} and determining the effect on the correction to the parton jet $p_T$ (dijet $M_{inv}$). For the inclusive jet (dijet) sample this uncertainty is in the range of $0.11-0.18\,\mathrm{GeV}/c$  ($0.15-0.32\, \mathrm{GeV}/c^2$).

Other leading systematic uncertainties on the jet $p_T$ (dijet $M_{inv}$) are associated with the detector response. These include the uncertainty on how well the calorimeter response to hadrons is modeled in GEANT. This uncertainty ranges from $0.8\%$ ($0.9\%$) at low $p_T$ ($M_{inv}$) to  $1.1\%$ ($1.1\%$) at high $p_T$ ($M_{inv}$). The detector response uncertainties also include how well the calorimeter gains are determined. The uncertainty on the gain calibration was estimated to be $3.2\%$, contributing to an uncertainty on jet $p_T$ (dijet $M_{inv}$) that ranges from $2.0\%$ ($1.6\%$) at low $p_T$ ($M_{inv}$) to $1.3\%$ ($1.2\%$) at high $p_T$  ($M_{inv}$).
Tracking inefficiency, conservatively estimated to be 4\%, is a further source of uncertainty. The combined uncertainty on jet $p_T$ (dijet $M_{inv}$) increases with $p_T$ ($M_{inv}$) from $0.19\,\mathrm{GeV}/c$ ($0.53\,\mathrm{GeV}/c^2$) to $0.62\,\mathrm{GeV}/c$ ($1.31\,\mathrm{GeV}/c^2$).

The $A_{LL}$ values for inclusive jets with $0.5<|\eta|<1$ and $|\eta|<0.5$ are reported in Table \ref{tab:IncJetAsymSysTable} together with the statistical and the total of the aforementioned, as well as smaller, systematic uncertainties. Table \ref{tab:DijetAsymSysTable} contains the dijet results and their uncertainties for the $\textrm{sign}(\eta_1) = \textrm{sign}(\eta_2)$ and $\textrm{sign}(\eta_1) \neq \textrm{sign}(\eta_2)$ topologies. The relative luminosity and beam polarization uncertainties are common to all data and are reported separately. The parity nonconserving longitudinal single-spin asymmetries for jets and dijets were found to vanish to within their statistical uncertainties, in agreement with expectations, and thus provide no evidence for substantial unaccounted systematic effects.

\begin{table}[th!]
\begin{tabular}{ccc}
\hline\hline

Jet $\eta$ & \begin{tabular}{@{}c@{}}{$p_T$ $\pm$ (Sys)} \\ {[GeV/$c$]}\end{tabular}   & $A_{LL} \pm$ (Stat) $\pm$ (Sys)  \\ \hline
	&	{ }6.15	$\pm$	0.19	&	-0.0002	$\pm$	0.0017	$\pm$	0.0004	\\
	&	{ }7.34	$\pm$	0.20	&	{ }0.0009	$\pm$	0.0019	$\pm$	0.0004	\\
	&	{ }9.50	$\pm$	0.28	&	{ }0.0008	$\pm$	0.0012	$\pm$	0.0004	\\
	&	11.34	$\pm$	0.28	&	{ }0.0033	$\pm$	0.0014	$\pm$	0.0004	\\
	&	13.25	$\pm$	0.31	&	{ }0.0024	$\pm$	0.0018	$\pm$	0.0004	\\
$0.5 < |\eta| < 1$	&	15.47	$\pm$	0.36	&	{ }0.0021	$\pm$	0.0025	$\pm$	0.0004	\\
	&	18.07	$\pm$	0.37	&	{ }0.0114	$\pm$	0.0037	$\pm$	0.0005	\\
	&	21.16	$\pm$	0.41	&	{ }0.0123	$\pm$	0.0058	$\pm$	0.0006	\\
	&	24.68	$\pm$	0.48	&	{ }0.0206	$\pm$	0.0099	$\pm$	0.0010	\\
	&	28.56	$\pm$	0.53	&	{ }0.0531	$\pm$	0.0180	$\pm$	0.0013	\\
	&	32.90	$\pm$	0.60	&	{ }0.0232	$\pm$	0.0357	$\pm$	0.0016	\\ \hline
	&	{ }6.00	$\pm$	0.20	&	{ }0.0002	$\pm$	0.0014	$\pm$	0.0015	\\ 
	&	{ }7.31	$\pm$	0.22	&	{ }0.0006	$\pm$	0.0017	$\pm$	0.0012	\\
	&	{ }9.47	$\pm$	0.25	&	{ }0.0034	$\pm$	0.0010	$\pm$	0.0011	\\
	&	11.33	$\pm$	0.31	&	{ }0.0041	$\pm$	0.0011	$\pm$	0.0009	\\
	&	13.36	$\pm$	0.34	&	{ }0.0070	$\pm$	0.0014	$\pm$	0.0008	\\
$|\eta| < 0.5 $	&	15.74	$\pm$	0.35	&	{ }0.0045	$\pm$	0.0019	$\pm$	0.0007	\\
	&	18.43	$\pm$	0.38	&	{ }0.0182	$\pm$	0.0028	$\pm$	0.0008	\\
	&	21.49	$\pm$	0.44	&	{ }0.0220	$\pm$	0.0043	$\pm$	0.0009	\\
	&	25.10	$\pm$	0.51	&	{ }0.0196	$\pm$	0.0072	$\pm$	0.0011	\\
	&	29.17	$\pm$	0.56	&	{ }0.0348	$\pm$	0.0127	$\pm$	0.0014	\\
	&	33.81	$\pm$	0.62	&	{ }0.0515	$\pm$	0.0239	$\pm$	0.0015	\\
\hline\hline
\end{tabular}
\caption{Parton inclusive-jet $p_T$ and $A_{LL}$ values with associated uncertainties for jet-$\eta$ regions $0.5<|\eta|<1$ and $|\eta|<0.5$. The $A_{LL}$ uncertainty contribution of $0.0007$ from uncertainty in the relative luminosity measurement and $6.1\%$ from the beam polarization uncertainty are common to all data points. They are in addition to the listed systematic uncertainty values.}
\label{tab:IncJetAsymSysTable}
\end{table}

\begin{table}[th!]
\begin{tabular}{ccc}
\hline\hline
Topology & \begin{tabular}{@{}c@{}}{$M_{inv}$ $\pm$ (Sys)} \\ {[GeV/$c^2$]}\end{tabular} & $A_{LL} \pm$ (Stat) $\pm$ (Sys)  \\ \hline
                                                   & 20.29 $\pm$ 0.53 & 0.0071 $\pm$ 0.0036 $\pm$  0.0009 \\ 
                                                   & 23.50 $\pm$ 0.61 & 0.0049 $\pm$ 0.0028 $\pm$  0.0007 \\ 
                                                   & 28.28 $\pm$ 0.66 & 0.0017 $\pm$ 0.0035 $\pm$  0.0008 \\ 
    $\textrm{Sign}(\eta_1) = \textrm{sign}(\eta_2)$& 34.15 $\pm$ 0.78 & 0.0137 $\pm$ 0.0051 $\pm$  0.0009 \\ 
                                                   & 40.96 $\pm$ 0.89 & 0.0316 $\pm$ 0.0081 $\pm$  0.0011 \\ 
                                                   & 50.75 $\pm$ 1.03 & 0.0232 $\pm$ 0.0121 $\pm$  0.0015 \\ 
                                                   & 69.11 $\pm$ 1.31 & 0.0228 $\pm$ 0.0418 $\pm$  0.0033 \\ \hline
                                                   & 20.48 $\pm$ 0.62 & 0.0067 $\pm$ 0.0040 $\pm$  0.0005 \\ 
                                                   & 23.65 $\pm$ 0.59 & 0.0024 $\pm$ 0.0027 $\pm$  0.0005 \\ 
                                                   & 28.50 $\pm$ 0.69 & 0.0052 $\pm$ 0.0032 $\pm$  0.0006 \\ 
$\textrm{Sign}(\eta_1) \neq \textrm{sign}(\eta_2)$ & 34.38 $\pm$ 0.79 & 0.0110 $\pm$ 0.0044 $\pm$  0.0007 \\ 
                                                   & 41.38 $\pm$ 0.92 & 0.0201 $\pm$ 0.0068 $\pm$  0.0009 \\ 
                                                   & 51.25 $\pm$ 1.08 & 0.0240 $\pm$ 0.0097 $\pm$  0.0012 \\ 
                                                   & 69.96 $\pm$ 1.31 & 0.0934 $\pm$ 0.0304 $\pm$  0.0020 \\ \hline\hline
\end{tabular}
\caption{Parton dijet invariant mass and $A_{LL}$ for the $\textrm{sign}(\eta_1) = \textrm{sign}(\eta_2)$ and $\textrm{sign}(\eta_1) \neq \textrm{sign}(\eta_2)$ topologies. The $A_{LL}$ uncertainty contribution of $0.0007$ from uncertainty in the relative luminosity measurement and $6.1\%$ from the beam polarization uncertainty are common to all data points. They are in addition to the listed systematic uncertainty values.}
\label{tab:DijetAsymSysTable}
\end{table}

\begin{figure}[th!]
\includegraphics[width=\columnwidth]{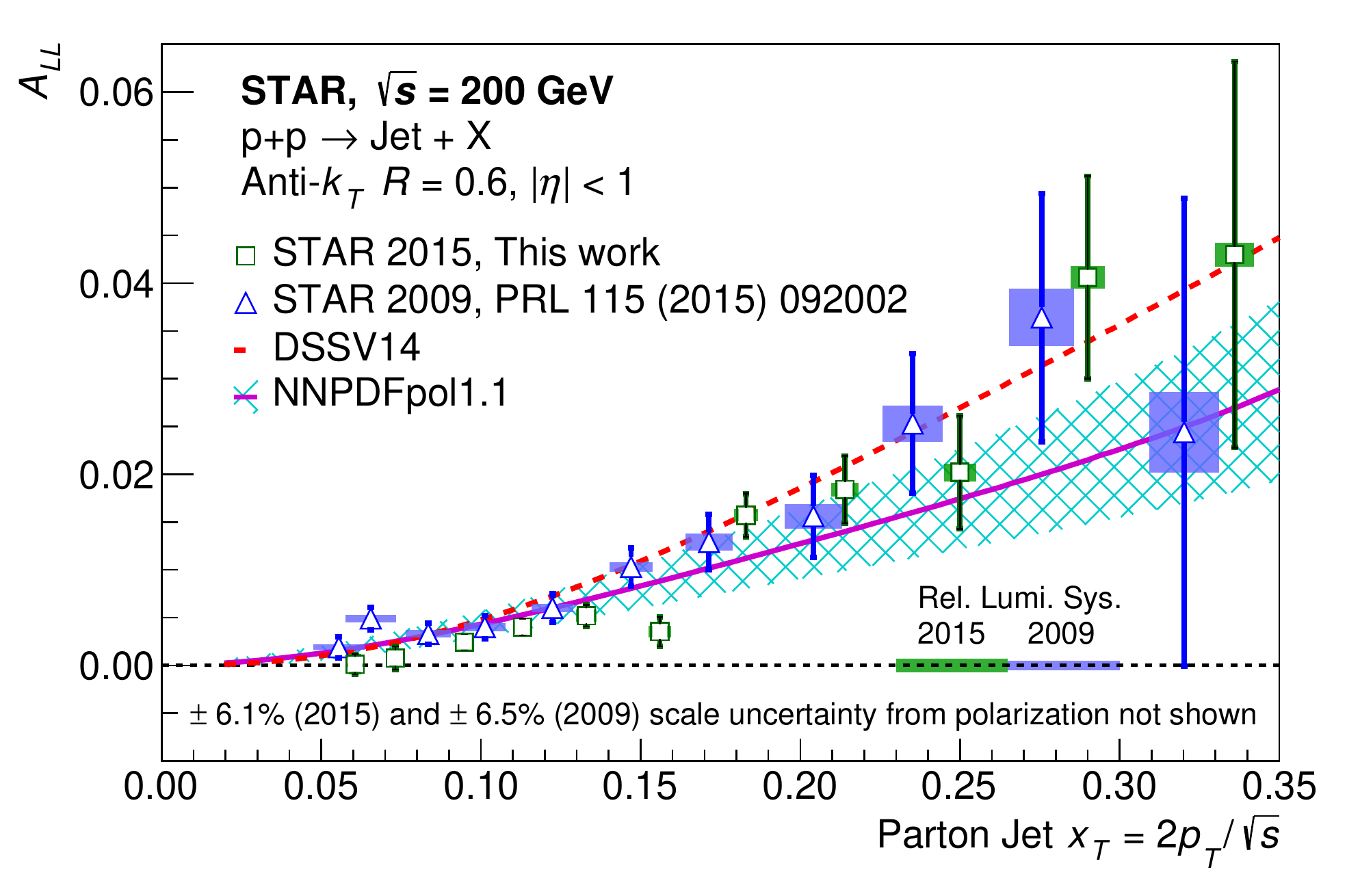}
\caption{\label{fig:IncJetALL} $A_{LL}$ for inclusive jets with $|\eta|<1.0$ versus $x_T$. The square markers show the present data, whereas the triangle markers show the data of Ref.~\cite{Adamczyk:2014ozi}. The error bars show the size of the statistical uncertainties, whereas the boxes indicate the size of the point-to-point systematic uncertainties in $A_{LL}$ (vertical extent) and $x_T$ (horizontal extent). The bands on the horizontal axis represent the relative luminosity uncertainty, which are common to all data points. The curves show the expected $A_{LL}$ values for the DSSV14~\cite{deFlorian:2014yva}
and NNPDFpol1.1~\cite{Nocera:2014gqa} parton distributions. The hatched area indicates the size of the uncertainty in the NNPDFpol1.1 expectation.}
\end{figure}

\begin{figure}[th!]
\includegraphics[width=\columnwidth]{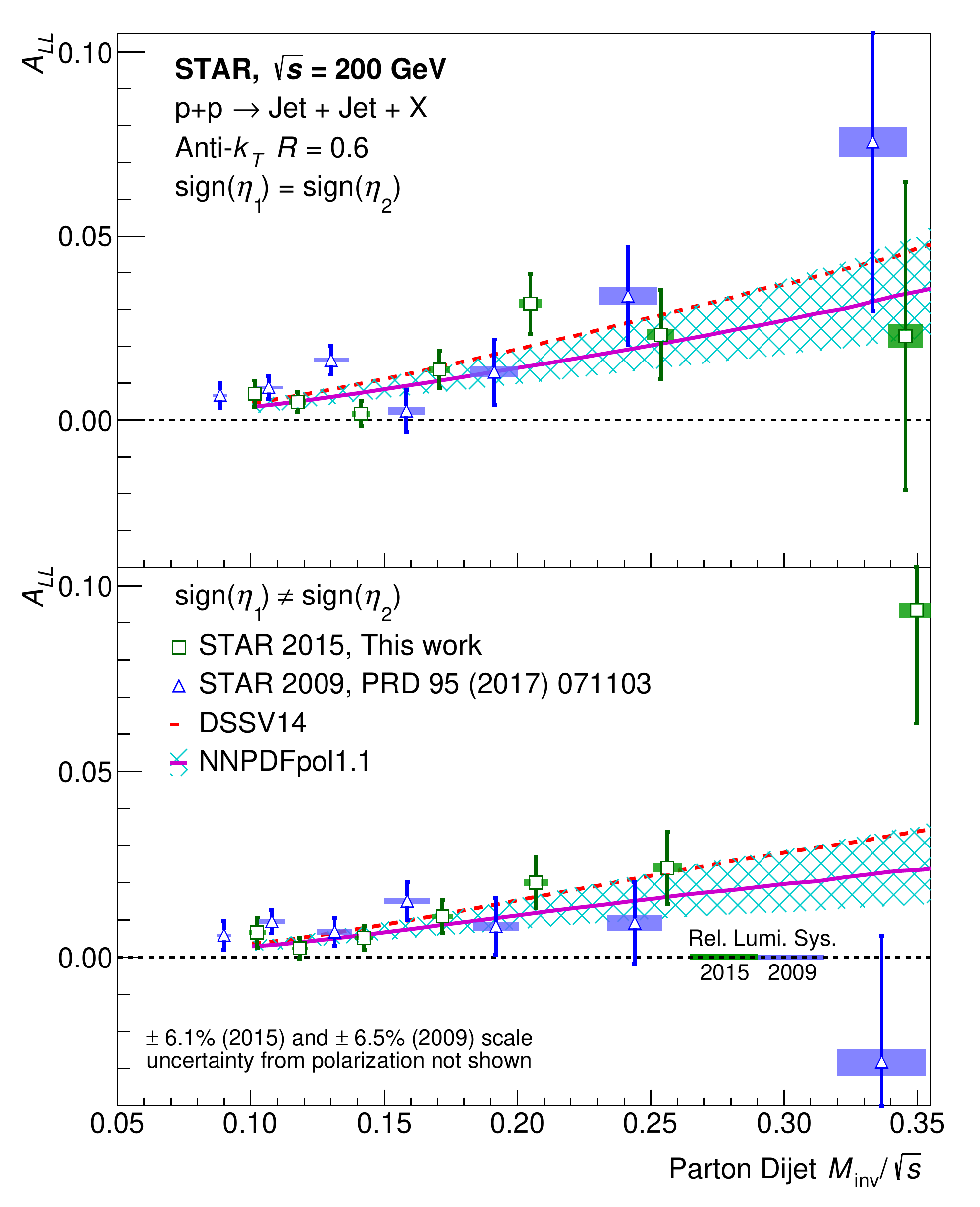}
\caption{\label{fig:DiJetALL}
$A_{LL}$ versus $M_{inv}/\sqrt{s}$ for dijets with the $\textrm{sign}(\eta_1) = \textrm{sign}(\eta_2)$ (top) and $\textrm{sign}(\eta_1) \neq \textrm{sign}(\eta_2)$ (bottom) event topologies. The square markers show the present data, whereas the triangle markers show the data of Ref.~\cite{Adamczyk:2016okk}. The results are compared to theoretical predictions for dijets from DSSV14~\cite{deFlorian:2014yva} and NNPDFpol1.1~\cite{Nocera:2014gqa} with its uncertainty.}
\end{figure}

About 98\% of the dijet events contain at least one jet that satisfies the inclusive jet requirements. This leads to statistical correlations of up to 0.27 between the inclusive jet and dijet $A_{LL}$. The 4\% of the inclusive jet event sample with two jets that both satisfy the inclusive jet requirements introduce a correlation among the inclusive jet $A_{LL}$ that ranges from about 0.005 at low $p_T$ to 0.06 at high $p_T$. There are no such statistical correlations among the dijet $A_{LL}$. Point-to-point correlations also originate from the underlying-event systematic uncertainties for $A_{LL}$ and the polarized PDF uncertainties in the evaluation of trigger bias. Systematic uncertainties dominate in the correlations at low $p_T$, estimated to be $< 0.07$, with exception of the correlations between low-$p_T$ intervals for inclusive jets with $|\eta|<0.5$. The large correlations up to 0.55 in this region originate from dominating underlying-event uncertainties which are fully correlated between intervals. At high $p_T$, the dominant effects are statistical in origin. The Supplemental Material contains detailed correlation matrices for all measurements \cite{SupplementalMaterial}.

Figures~\ref{fig:IncJetALL} and \ref{fig:DiJetALL} show the inclusive jet and dijet $A_{LL}$, respectively, versus $x_T = 2p_T/\sqrt{s}$ and $M_{inv}/\sqrt{s}$ corrected to the parton level. The data are plotted alongside our prior $\sqrt{s} = 200\,\mathrm{GeV}$ data~\cite{Adamczyk:2014ozi,Adamczyk:2016okk}. The $x_T$ and $M_{inv}/\sqrt{s}$ values for the present data are slightly higher and have smaller uncertainties than those for our prior data mainly because of refined treatment of the underlying-event correction. Also shown are theory expectations from the DSSV14~\cite{deFlorian:2014yva} (red dashed curve) and NNPDFpol1.1~\cite{Nocera:2014gqa} (purple continuous curve) global analyses. NNPDFpol1.1 has 100 publicly available and equally probable replicas and their root mean square corresponding to the  one-sigma error band is represented by the cyan hatched region.

The present and prior data are in good agreement. The prior inclusive jet data \cite{Adamczyk:2014ozi} are included in the theory expectations and provide stringent constraints on $\Delta g(x,Q^2)$ for $x > 0.05$. The theory expectations are in good agreement with the present data, which have further improved accuracy.

In summary, we have presented $A_{LL}$ for midrapidity inclusive jet and dijet production based on data recorded by the STAR Collaboration during RHIC operations with polarized proton beams at $\sqrt{s}=200\,\mathrm{GeV}$ in 2015. The data provide sensitivity to the polarized gluon distribution  $\Delta g(x,Q^2)$ for gluon momentum fractions $0.05 \lesssim x \lesssim 0.5$. Our new results are consistent with and have better precision than our prior data at $\sqrt{s}=200\,\mathrm{GeV}$. The results provide further evidence that $\Delta g(x,Q^2)$ is positive for $x > 0.05$.

We thank the RHIC Operations Group and RCF at BNL, the NERSC Center at LBNL, and the Open Science Grid consortium for providing resources and support. This work was supported in part by the Office of Nuclear Physics within the U.S. DOE Office of Science, the U.S. National Science Foundation, the Ministry of Education and Science of the Russian Federation, National Natural Science Foundation of China, Chinese Academy of Science, the Ministry of Science and Technology of China and the Chinese Ministry of Education, the Higher Education Sprout Project by Ministry of Education at NCKU, the National Research Foundation of Korea, Czech Science Foundation and Ministry of Education, Youth and Sports of the Czech Republic, Hungarian National Research, Development and Innovation Office, New National Excellency Programme of the Hungarian Ministry of Human Capacities, Department of Atomic Energy and Department of Science and Technology of the Government of India, the National Science Centre of Poland, the Ministry of Science, Education and Sports of the Republic of Croatia, RosAtom of Russia and German Bundesministerium fur Bildung, Wissenschaft, Forschung and Technologie, Helmholtz Association, Ministry of Education, Culture, Sports, Science, and Technology and Japan Society for the Promotion of Science.

\newpage

\bibliographystyle{apsrev4-1}
\bibliography{jet200GeV2015Bib}

\newpage
\onecolumngrid
\appendix
\section*{Supplemental Material}

\begin{table}[ht!]
\begin{tabular}{cccccccccccc}
\hline\hline
$p_{T}$ $[\mathrm{GeV}/c]$ &6.15&7.34 &9.50 &11.34 &13.25 &15.47 &18.07 &21.16 &24.68 &28.56 &32.90\\ \hline
6.15& 1&0.043 &0.062 &0.051 &0.037 &0.026 &0.019 &0.012 &0.011 &0.007 &0.003 \\
7.34&  &1 &0.051 &0.045 &0.033 &0.023 &0.018 &0.012 &0.011 &0.007 &0.003 \\
9.50&  & &1&0.069 &0.051 &0.037 &0.029 &0.019 &0.017 &0.011 &0.005 \\
11.34&  & & &1&0.051 &0.039 &0.032 &0.020 &0.019 &0.013 &0.006 \\
13.25&  & & & &1&0.034 &0.028 &0.019 &0.017 &0.010 &0.005 \\
15.47&  & & & & &1&0.027 &0.019 &0.016 &0.010 &0.004 \\
18.07&  & & & & & &1&0.023 &0.019 &0.011 &0.005 \\
21.16&  & & & & & & &1&0.022 &0.014 &0.006 \\
24.68&  & & & & & & & &1&0.020 &0.011 \\
28.56&  & & & & & & & & &1&0.016 \\
32.90&  & & & & & & & & & &1\\
\hline\hline
\end{tabular}
\caption{The correlation matrix for the point-to-point uncertainties in the inclusive jet measurements for jets in the $\eta$ range $0.5<|\eta|<1$. At low $p_T$, the dominant effects arise from correlated systematic uncertainties, whereas at high $p_T$, the dominant effects arise from the statistical correlations when two jets in the same event satisfy all the inclusive jet cuts. The $A_{LL}$ uncertainty contribution of $0.0007$ from uncertainty in the relative luminosity measurement and $6.1\%$ from the beam polarization uncertainty, which are common to all the data points, are separated from the listed values.}
\label{tab:MatrixIncJetFwdIncJetFwd}
\end{table}

\begin{table}[ht!]
\begin{tabular}{cccccccccccc}
\hline\hline
$p_{T}$ $[\mathrm{GeV}/c]$& 6.00&7.31&9.47&11.33&13.36&15.74&18.43&21.49&25.10&29.17&33.81\\ \hline
6.00&1&0.448 &0.549 &0.463 &0.353 &0.244 &0.163 &0.102 &0.064 &0.035 &0.018 \\
7.31& &1&0.450 &0.379 &0.287 &0.198 &0.130 &0.080 &0.049 &0.027 &0.014 \\
9.47& & &1&0.470 &0.361 &0.251 &0.171 &0.109 &0.069 &0.038 &0.020 \\
11.33& & & &1&0.309 &0.217 &0.149 &0.095 &0.060 &0.033 &0.017 \\
13.36& & & & &1&0.177 &0.129 &0.085 &0.056 &0.031 &0.016 \\
15.74& & & & & &1&0.104 &0.072 &0.049 &0.027 &0.014 \\
18.43& & & & & & &1&0.075 &0.056 &0.032 &0.017 \\
21.49& & & & & & & &1&0.062 &0.040 &0.021 \\
25.10& & & & & & & & &1&0.056 &0.033 \\
29.17& & & & & & & & & &1&0.053 \\
33.81& & & & & & & & & & &1\\

\hline\hline
\end{tabular}
\caption{The correlation matrix for the point-to-point uncertainties in the inclusive jet measurements for jets in the $\eta$ range $|\eta|<0.5$. At low $p_T$, the dominant effects arise from correlated systematic uncertainties. Large correlations at this region originate from dominating underlying-event uncertainties which are fully correlated among the data points. At high $p_T$, the dominant effects arise from the statistical correlations when two jets in the same event satisfy all the inclusive jet cuts. The $A_{LL}$ uncertainty contribution of $0.0007$ from uncertainty in the relative luminosity measurement and $6.1\%$ from the beam polarization uncertainty, which are common to all the data points, are separated from the listed values.}
\label{tab:MatrixIncJetMidIncJetMid}
\end{table}

\begin{table}[ht!]
\begin{tabular}{cccccccccccc}
\hline\hline
$p_{T}$ $[\mathrm{GeV}/c]$&6.06&7.32&9.48&11.33&13.32&15.64&18.30&21.38&24.97&28.98&33.55\\ \hline
6.06&1&0.415 &0.516 &0.431 &0.325 &0.224 &0.148 &0.093 &0.058 &0.033 &0.017 \\
7.32& &1&0.422 &0.353 &0.267 &0.185 &0.123 &0.077 &0.048 &0.027 &0.014 \\
9.48& & &1&0.451 &0.346 &0.243 &0.166 &0.107 &0.068 &0.039 &0.020 \\
11.33& & & &1&0.303 &0.217 &0.153 &0.100 &0.066 &0.038 &0.020 \\
13.32& & & & &1&0.183 &0.136 &0.092 &0.062 &0.036 &0.018 \\
15.64& & & & & &1&0.117 &0.084 &0.057 &0.033 &0.016 \\
18.30& & & & & & &1&0.090 &0.066 &0.039 &0.019 \\
21.38& & & & & & & &1&0.078 &0.051 &0.025 \\
24.97& & & & & & & & &1&0.071 &0.042 \\
28.98& & & & & & & & & &1&0.066 \\
33.55& & & & & & & & & & &1\\

\hline\hline
\end{tabular}
\caption{The correlation matrix for the point-to-point uncertainties in the inclusive jet measurements for jets in the $\eta$ range $|\eta|<1$. At low $p_T$, the dominant effects arise from correlated systematic uncertainties. Large correlations at this region originate from dominating underlying-event uncertainties which are fully correlated among the data points. At high $p_T$, the dominant effects arise from the statistical correlations when two jets in the same event satisfy all the inclusive jet cuts. The $A_{LL}$ uncertainty contribution of $0.0007$ from uncertainty in the relative luminosity measurement and $6.1\%$ from the beam polarization uncertainty, which are common to all the data points, are separated from the listed values.}
\label{tab:MatrixIncJetFullIncJetFull}
\end{table}

\begin{table}[ht!]
\begin{tabular}{cccccccccccc}
\hline\hline
$p_{T}$ $[\mathrm{GeV}/c]$&6.15&7.34&9.50&11.34&13.25&15.47&18.07&21.16&24.68&28.56&32.90\\\hline
6.00&0.019 &0.021 &0.029 &0.033 &0.026 &0.020 &0.018 &0.012 &0.013 &0.009 &0.005 \\
7.31&0.013 &0.015 &0.019 &0.021 &0.017 &0.013 &0.012 &0.008 &0.008 &0.006 &0.003 \\
9.47&0.021 &0.024 &0.037 &0.042 &0.034 &0.027 &0.024 &0.016 &0.017 &0.011 &0.005 \\
11.33&0.019 &0.021 &0.034 &0.039 &0.033 &0.027 &0.024 &0.016 &0.016 &0.010 &0.005 \\
13.36&0.020 &0.022 &0.037 &0.044 &0.038 &0.034 &0.030 &0.021 &0.019 &0.012 &0.006 \\
15.74&0.016 &0.018 &0.030 &0.037 &0.035 &0.033 &0.033 &0.025 &0.021 &0.013 &0.006 \\
18.43&0.019 &0.021 &0.034 &0.041 &0.038 &0.038 &0.041 &0.035 &0.030 &0.019 &0.009 \\
21.49&0.015 &0.017 &0.027 &0.033 &0.030 &0.031 &0.037 &0.038 &0.038 &0.026 &0.013 \\
25.10&0.014 &0.015 &0.024 &0.028 &0.024 &0.024 &0.029 &0.035 &0.042 &0.037 &0.023 \\
29.17&0.009 &0.009 &0.015 &0.017 &0.014 &0.014 &0.017 &0.022 &0.032 &0.040 &0.030 \\
33.81&0.005 &0.005 &0.008 &0.009 &0.008 &0.007 &0.008 &0.010 &0.019 &0.029 &0.035 \\

\hline\hline
\end{tabular}
\caption{The correlation matrix for the point-to-point uncertainties coupling the inclusive jet measurements for jets in the $|\eta|<0.5$ (rows) and $0.5<|\eta|<1$ (columns) ranges. The $A_{LL}$ uncertainty contribution of $0.0007$ from uncertainty in the relative luminosity measurement and $6.1\%$ from the beam polarization uncertainty, which are common to all the data points, are separated from the listed values.}
\label{tab:MatrixIncJetFwdIncJetMid}
\end{table}

\begin{table}[ht!]
\begin{tabular}{cccccccc}
\hline\hline
\diagbox{\raisebox{1mm}{$p_{T}$ $[\mathrm{GeV}/c]$}}{\raisebox{-1mm}{$M_{inv}$ $[\mathrm{GeV}/c^2]$}}&20.29&23.50&28.28&34.15&40.96&50.75&69.11\\\hline
6.15&0.013 &0.013 &0.012 &0.008 &0.007 &0.009 &0.006 \\
7.34&0.021 &0.018 &0.014 &0.009 &0.008 &0.010 &0.007 \\
9.50&0.060 &0.056 &0.032 &0.015 &0.012 &0.016 &0.010 \\
11.34&0.063 &0.081 &0.060 &0.024 &0.014 &0.018 &0.012 \\
13.25&0.030 &0.074 &0.083 &0.045 &0.016 &0.014 &0.009 \\
15.47&0.011 &0.043 &0.076 &0.080 &0.036 &0.014 &0.007 \\
18.07&0.010 &0.020 &0.046 &0.078 &0.081 &0.030 &0.007 \\
21.16&0.006 &0.009 &0.020 &0.041 &0.085 &0.080 &0.005 \\
24.68&0.007 &0.009 &0.012 &0.017 &0.042 &0.114 &0.017 \\
28.56&0.005 &0.007 &0.007 &0.007 &0.015 &0.081 &0.068 \\
32.90&0.002 &0.003 &0.003 &0.003 &0.004 &0.027 &0.108 \\

\hline\hline
\end{tabular}
\caption{The correlation matrix for the point-to-point uncertainties coupling the inclusive jet measurement for jets in the $0.5<|\eta|<1$ region (rows) with dijet measurements with $\textrm{sign}(\eta_1) = \textrm{sign}(\eta_2)$ topology (columns). The $A_{LL}$ uncertainty contribution of $0.0007$ from uncertainty in the relative luminosity measurement and $6.1\%$ from the beam polarization uncertainty, which are common to all the data points, are separated from the listed values.}
\label{tab:MatrixDijetSameIncJetFwd}
\end{table}

\begin{table}[ht!]
\begin{tabular}{cccccccc}
\hline\hline
\diagbox{\raisebox{1mm}{$p_{T}$ $[\mathrm{GeV}/c]$}}{\raisebox{-1mm}{$M_{inv}$ $[\mathrm{GeV}/c^2]$}}&20.48&23.65&28.50&34.38&41.38&51.25&69.96\\ \hline
6.15&0.016 &0.015 &0.013 &0.008 &0.007 &0.009 &0.005 \\
7.34&0.020 &0.022 &0.016 &0.009 &0.008 &0.010 &0.005 \\
9.50&0.056 &0.064 &0.042 &0.019 &0.012 &0.016 &0.008 \\
11.34&0.035 &0.076 &0.071 &0.035 &0.016 &0.018 &0.009 \\
13.25&0.015 &0.052 &0.079 &0.063 &0.027 &0.015 &0.007 \\
15.47&0.011 &0.024 &0.060 &0.080 &0.061 &0.023 &0.005 \\
18.07&0.010 &0.013 &0.033 &0.059 &0.088 &0.060 &0.006 \\
21.16&0.006 &0.008 &0.014 &0.027 &0.063 &0.112 &0.011 \\
24.68&0.007 &0.009 &0.010 &0.013 &0.026 &0.114 &0.056 \\
28.56&0.005 &0.006 &0.007 &0.006 &0.009 &0.057 &0.130 \\
32.90&0.002 &0.003 &0.003 &0.002 &0.003 &0.018 &0.122 \\
\hline\hline
\end{tabular}
\caption{The correlation matrix for the point-to-point uncertainties coupling the inclusive jet measurement for jets in the $0.5<|\eta|<1$ region (rows) with dijet measurements with $\textrm{sign}(\eta_1) \neq \textrm{sign}(\eta_2)$ topology (columns). The $A_{LL}$ uncertainty contribution of $0.0007$ from uncertainty in the relative luminosity measurement and $6.1\%$ from the beam polarization uncertainty, which are common to all the data points, are separated from the listed values.}
\label{tab:MatrixDijetOppIncJetFwd}
\end{table}

\begin{table}[ht!]
\begin{tabular}{cccccccc}
\hline\hline
\diagbox{\raisebox{1mm}{$p_{T}$ $[\mathrm{GeV}/c]$}}{\raisebox{-1mm}{$M_{inv}$ $[\mathrm{GeV}/c^2]$}}&20.29&23.50&28.28&34.15&40.96&50.75&69.11\\ \hline
6.00&0.019 &0.020 &0.020 &0.013 &0.011 &0.015 &0.010 \\
7.31&0.024 &0.019 &0.013 &0.008 &0.007 &0.009 &0.006 \\
9.47&0.066 &0.062 &0.036 &0.017 &0.014 &0.018 &0.012 \\
11.33&0.077 &0.098 &0.070 &0.025 &0.013 &0.015 &0.010 \\
13.36&0.042 &0.106 &0.122 &0.068 &0.021 &0.017 &0.011 \\
15.74&0.014 &0.066 &0.120 &0.135 &0.061 &0.019 &0.009 \\
18.43&0.016 &0.033 &0.078 &0.138 &0.149 &0.055 &0.011 \\
21.49&0.013 &0.018 &0.038 &0.075 &0.160 &0.161 &0.011 \\
25.10&0.012 &0.015 &0.020 &0.030 &0.078 &0.236 &0.038 \\
29.17&0.007 &0.009 &0.010 &0.011 &0.024 &0.164 &0.168 \\
33.81&0.004 &0.005 &0.006 &0.005 &0.007 &0.057 &0.259 \\
\hline\hline
\end{tabular}
\caption{The correlation matrix for the point-to-point uncertainties coupling the inclusive jet measurement for jets in the $|\eta|<0.5$ region (rows) with dijet measurements with $\textrm{sign}(\eta_1) = \textrm{sign}(\eta_2)$ topology (columns). The $A_{LL}$ uncertainty contribution of $0.0007$ from uncertainty in the relative luminosity measurement and $6.1\%$ from the beam polarization uncertainty, which are common to all the data points, are separated from the listed values.}
\label{tab:MatrixDijetSameIncJetMid}
\end{table}

\begin{table}[ht!]
\begin{tabular}{cccccccc}
\hline\hline
\diagbox{\raisebox{1mm}{$p_{T}$ $[\mathrm{GeV}/c]$}}{\raisebox{-1mm}{$M_{inv}$ $[\mathrm{GeV}/c^2]$}}&20.48&23.65&28.50&34.38&41.38&51.25&69.96\\ \hline
6.00&0.022 &0.021 &0.020 &0.013 &0.011 &0.015 &0.007 \\
7.31&0.027 &0.022 &0.014 &0.008 &0.007 &0.009 &0.004 \\
9.47&0.075 &0.070 &0.042 &0.019 &0.013 &0.018 &0.009 \\
11.33&0.070 &0.103 &0.079 &0.032 &0.014 &0.016 &0.008 \\
13.36&0.031 &0.099 &0.124 &0.082 &0.028 &0.018 &0.008 \\
15.74&0.013 &0.053 &0.111 &0.138 &0.080 &0.024 &0.007 \\
18.43&0.016 &0.027 &0.069 &0.122 &0.156 &0.078 &0.009 \\
21.49&0.013 &0.017 &0.033 &0.063 &0.139 &0.183 &0.013 \\
25.10&0.012 &0.015 &0.019 &0.027 &0.063 &0.226 &0.062 \\
29.17&0.007 &0.009 &0.010 &0.010 &0.020 &0.138 &0.196 \\
33.81&0.004 &0.005 &0.006 &0.004 &0.007 &0.046 &0.248 \\
\hline\hline
\end{tabular}
\caption{The correlation matrix for the point-to-point uncertainties coupling the inclusive jet measurement for jets in the $|\eta|<0.5$ region (rows) with dijet measurements with $\textrm{sign}(\eta_1) \neq \textrm{sign}(\eta_2)$ topology (columns). The $A_{LL}$ uncertainty contribution of $0.0007$ from uncertainty in the relative luminosity measurement and $6.1\%$ from the beam polarization uncertainty, which are common to all the data points, are separated from the listed values.}
\label{tab:MatrixDijetOppIncJetMid}
\end{table}

\begin{table}[ht!]
\begin{tabular}{cccccccc}
\hline\hline
\diagbox{\raisebox{1mm}{$p_{T}$ $[\mathrm{GeV}/c]$}}{\raisebox{-1mm}{$M_{inv}$ $[\mathrm{GeV}/c^2]$}}&20.29&23.50&28.28&34.15&40.96&50.75&69.11\\\hline
6.06&0.019 &0.019 &0.019 &0.012 &0.011 &0.014 &0.009 \\
7.32&0.029 &0.024 &0.017 &0.011 &0.009 &0.012 &0.008 \\
9.48&0.078 &0.073 &0.042 &0.020 &0.016 &0.020 &0.014 \\
11.33&0.093 &0.119 &0.086 &0.032 &0.018 &0.022 &0.014 \\
13.32&0.050 &0.125 &0.142 &0.079 &0.026 &0.022 &0.014 \\
15.64&0.018 &0.077 &0.139 &0.153 &0.069 &0.023 &0.012 \\
18.30&0.017 &0.036 &0.087 &0.154 &0.164 &0.060 &0.012 \\
21.38&0.014 &0.019 &0.042 &0.083 &0.176 &0.173 &0.012 \\
24.97&0.012 &0.015 &0.021 &0.033 &0.085 &0.253 &0.039 \\
28.98&0.008 &0.010 &0.011 &0.012 &0.027 &0.178 &0.174 \\
33.55&0.004 &0.005 &0.006 &0.005 &0.008 &0.061 &0.272 \\
\hline\hline
\end{tabular}
\caption{The correlation matrix for the point-to-point uncertainties coupling the inclusive jet measurement for jets in the $|\eta|<1$ region (rows) with dijet measurements with $\textrm{sign}(\eta_1) = \textrm{sign}(\eta_2)$ topology (columns). The $A_{LL}$ uncertainty contribution of $0.0007$ from uncertainty in the relative luminosity measurement and $6.1\%$ from the beam polarization uncertainty, which are common to all the data points, are separated from the listed values.}
\label{tab:MatrixDijetSameIncJetFull}
\end{table}

\begin{table}[ht!]
\begin{tabular}{cccccccc}
\hline\hline
\diagbox{\raisebox{1mm}{$p_{T}$ $[\mathrm{GeV}/c]$}}{\raisebox{-1mm}{$M_{inv}$ $[\mathrm{GeV}/c^2]$}}&20.48&23.65&28.50&34.38&41.38&51.25&69.96\\\hline
6.06&0.023 &0.020 &0.019 &0.013 &0.010 &0.014 &0.007 \\
7.32&0.031 &0.028 &0.019 &0.011 &0.009 &0.012 &0.006 \\
9.48&0.084 &0.083 &0.051 &0.023 &0.016 &0.020 &0.010 \\
11.33&0.073 &0.120 &0.099 &0.044 &0.019 &0.022 &0.011 \\
13.32&0.033 &0.107 &0.141 &0.100 &0.037 &0.023 &0.011 \\
15.64&0.017 &0.056 &0.123 &0.155 &0.098 &0.033 &0.009 \\
18.30&0.017 &0.028 &0.073 &0.130 &0.173 &0.095 &0.010 \\
21.38&0.014 &0.018 &0.034 &0.066 &0.147 &0.210 &0.017 \\
24.97&0.012 &0.015 &0.019 &0.027 &0.064 &0.245 &0.081 \\
28.98&0.008 &0.010 &0.011 &0.011 &0.021 &0.143 &0.231 \\
33.55&0.004 &0.005 &0.006 &0.004 &0.007 &0.047 &0.270 \\
\hline\hline
\end{tabular}
\caption{The correlation matrix for the point-to-point uncertainties coupling the inclusive jet measurement for jets in the $|\eta|<1$ region (rows) with dijet measurements with $\textrm{sign}(\eta_1) \neq \textrm{sign}(\eta_2)$ topology (columns). The $A_{LL}$ uncertainty contribution of $0.0007$ from uncertainty in the relative luminosity measurement and $6.1\%$ from the beam polarization uncertainty, which are common to all the data points, are separated from the listed values.}
\label{tab:MatrixDijetOppIncJetFull}
\end{table}

\begin{table}[ht!]
\begin{tabular}{cccccccc}
\hline\hline
$M_{inv}$ $[\mathrm{GeV}/c^2]$&20.29&23.50&28.28&34.15&40.96&50.75&69.11\\\hline
20.29&1&0.044 &0.033 &0.020 &0.013 &0.012 &0.006 \\
23.50& &1&0.038 &0.023 &0.016 &0.015 &0.008 \\
28.28& & &1&0.019 &0.013 &0.014 &0.008 \\
34.15& & & &1&0.009 &0.009 &0.005 \\
40.96& & & & &1&0.007 &0.004 \\
50.75& & & & & &1&0.006 \\
69.11& & & & & & &1\\
\hline\hline
\end{tabular}
\caption{The correlation matrix for the point-to-point uncertainties for dijet measurements with $\textrm{sign}(\eta_1) = \textrm{sign}(\eta_2)$ topology. The $A_{LL}$ uncertainty contribution of $0.0007$ from uncertainty in the relative luminosity measurement and $6.1\%$ from the beam polarization uncertainty, which are common to all the data points, are separated from the listed values.}
\label{tab:MatrixDijetSame}
\end{table}

\begin{table}[ht!]
\begin{tabular}{cccccccc}
\hline\hline
$M_{inv}$ $[\mathrm{GeV}/c^2]$&20.48&23.65&28.50&34.38&41.38&51.25&69.96\\\hline
20.48&1&0.011 &0.012 &0.008 &0.006 &0.008 &0.004 \\
23.65& &1&0.016 &0.010 &0.008 &0.010 &0.005 \\
28.50& & &1&0.011 &0.009 &0.011 &0.006 \\
34.38& & & &1&0.006 &0.008 &0.004 \\
41.38& & & & &1&0.006 &0.003 \\
51.25& & & & & &1&0.004 \\
69.96& & & & & & &1\\
\hline\hline
\end{tabular}
\caption{The correlation matrix for the point-to-point uncertainties for dijet measurements with $\textrm{sign}(\eta_1) \neq \textrm{sign}(\eta_2)$ topology. The $A_{LL}$ uncertainty contribution of $0.0007$ from uncertainty in the relative luminosity measurement and $6.1\%$ from the beam polarization uncertainty, which are common to all the data points, are separated from the listed values.}
\label{tab:MatrixDijetOpp}
\end{table}

\begin{table}[ht!]
\begin{tabular}{cccccccc}
\hline\hline
$M_{inv}$ $[\mathrm{GeV}/c^2]$&20.29&23.50&28.28&34.15&40.96&50.75&69.11\\\hline
20.48&0.008 &0.010 &0.010 &0.007 &0.006 &0.008 &0.005 \\
23.65&0.010 &0.013 &0.013 &0.009 &0.008 &0.010 &0.007 \\
28.50&0.011 &0.014 &0.015 &0.010 &0.008 &0.011 &0.007 \\
34.38&0.007 &0.009 &0.010 &0.007 &0.006 &0.007 &0.005 \\
41.38&0.006 &0.008 &0.008 &0.006 &0.005 &0.006 &0.004 \\
51.25&0.008 &0.010 &0.011 &0.007 &0.006 &0.008 &0.006 \\
69.96&0.004 &0.005 &0.005 &0.004 &0.003 &0.004 &0.003 \\
\hline\hline
\end{tabular}
\caption{The correlation matrix for the point-to-point uncertainties coupling dijet measurements with $\textrm{sign}(\eta_1) \neq \textrm{sign}(\eta_2)$ (rows) and $\textrm{sign}(\eta_1) = \textrm{sign}(\eta_2)$ (columns) topologies. The $A_{LL}$ uncertainty contribution of $0.0007$ from uncertainty in the relative luminosity measurement and $6.1\%$ from the beam polarization uncertainty, which are common to all the data points, are separated from the listed values.}
\label{tab:MatrixDijetOppSame}
\end{table}

\end{document}